khahn.tex000064400003230000024000000473040642056206600132070ustar00khahnusers00000000000000
\documentstyle[twocolumn,prb,aps]{revtex}
\input {psfig.tex}
\begin{document}
\twocolumn[\hsize\textwidth\columnwidth\hsize\csname @twocolumnfalse\endcsname
\draft

\begin{title}
{Pairing in the quantum Hall system}
\end{title}

\author{
    Kang-Hun Ahn and K. J. Chang
     }
\address{     
Department of Physics, Korea Advanced Institute of Science and Technology,\\
  373-1 Kusung-dong, Yusung-ku, Taejon, Korea 
          }
\maketitle

\date{\today}
\widetext

\begin{abstract}
We find an analogy between the single skyrmion state in the quantum
Hall system and the BCS superconducting state and address that the
quantum mechanical origin of the skyrmion is electronic pairing.
The skyrmion phase is found to be unstable for magnetic fields above the
critical field $B_{c}(T)$ at temperature $T$, which is well represented
by the relation $B_c(T)/B_{c}(0) \approx {[1-(T/T_c)^3]}^{1/2}$.
\vspace{1cm}

\end{abstract}
]

\narrowtext
\newpage

In the quantum Hall system, \cite{prange} skyrmions, i.e.,
charged spin textures, have been attracting much attention.
When Barret and his coworkers found that the electron spin polarization
drops precipitously on either side of $\nu=1$ ($\nu \equiv N/N_{\phi}$,
where N is the total number of electrons and $N_{\phi}$ is the Landau
level degeneracy), \cite{barret} the ground state was thought to consist
of skyrmion which was already theoretically studied through the nonlinear
sigma model \cite{lee} or the Hartree-Fock approach for zero temperature.
\cite{fertig}
Later, the skyrmion was described by a microscopic model without using
the sigma model, \cite{macdonald} however, electron-electron interactions
were assumed to be unrealistic point interactions.
From finite-size calculations for the electron-electron Coulomb interactions, 
Xie and He \cite{xie} found the lowest energy states, which are nearly
degenerated but have different total spins, in weak Zeeman coupling limit.
These low-lying states were used to describe skyrmions, which were
suggested to be symmetry-breaking states. \cite{xie,ahn}
However, it is still desirable to know the quantum mechanical origin
of skyrmions.
Here we find that the quamtum Hall system has a pairing effect, similar
to superconducting systems.
Since the BCS theory has been very successful in superconductivity, \cite{bcs}
we may use the same theoretical approach to understand the role of
electron-electron interactions in the quantum Hall system.

In this work, we present a picture that the skyrmion state consists
of electron-hole pairs, as the BCS superconducting state made of the Cooper
pairs.
Without the nonlinear sigma model, we are able to derive the skyrmion wave
function near the filling factor $\nu$ = 1, which is identical to that of
Fertig and his coworkers, \cite{fertig} from the eigenstates of the mean-field
pairing Hamiltonian. 
We find that the skyrmion phase is unstable for temperatures or magnetic
fields above their critical values.

Let us consider a quantum Hall system of $N$ electrons with the lowest
Landau level degeneracy $N_{\phi}=N+1$, which has the filling factor
($\nu = 1 - \epsilon$) slightly less than one.
The system with $\nu = 1 + \epsilon$ can be also analyzed using the
well known particle-hole symmetry.
The quantum Hall sample is assumed to have no impurities, so that
the skyrmion size is macroscopic in weak Zeeman coupling limit.
For sufficiently high magnetic fields, the ground state of the system will
be in a state $\Psi_{0}^{+}$ with all the spins aligned along the same up
direction and one hole at the origin.
When the applied magnetic field is reduced to lower the Zeeman energy,
$\Psi_0^{+}$ becomes unstable against the creation of an electron-hole
pair, which is described by $c^{\dagger}_{m,\downarrow}c_{m+1,\uparrow}
|\Psi_{0}^{+} \rangle$, where $c_{m,\sigma}^{\dagger}$ and $c_{m,\sigma}$
are the creation and annihilation operators for the electron in the
angular momentum $l = m$ and spin $\sigma$ state.
Then, the single pair states with different $m$'s can be linearly
combined to form the new ground state or the low-lying excited states
of the system. 

In analogy with the BCS pairing Hamiltonian, there exists an interaction
potential $V$ representing {\it pair-pair} correlations, which are responsible
for the rigidity of the pair condensate.
By noting that electron-electron interactions are purely the Coulomb
interactions in our theory, the potential energy $V$ in the Hamiltonian
can be written as
\begin{eqnarray}
V= \sum_{m \neq n} V_{m,n+1,n,m+1} c_{m,\downarrow}^{\dagger}
c_{n+1,\uparrow}^{\dagger} c_{m+1,\uparrow} c_{n,\downarrow},
\label{v}
\end{eqnarray}
where $V_{m_1,m_2,m_3,m_4}$ is the Coulomb interaction matrix between
the single particle states with $l = m_1, m_2, m_3$, and $m_4$ in the
2-dimensional system. \cite{interaction}
As in the case of superconductivity, the electron-hole pairings can lead to
a phase transition, which is called a {\it skyrmionic transition}
(this notion will be clearly discussed later).
In this transition, the topological charge $Q$ of the state changes
abruptly from -1 to 0 at the critical temperature $T_{c}$, which
depends on the strength of the Zeeman coupling.
Note that although the Coulomb interaction is purely repulsive, the
expectation value of $V$ can be negative, which makes the skyrmion phase stable.
Using the Hartree-Fock decomposition, one can easily show that  
the peculiar form of $V$ allows for the scattering between the states
with different total spins ($S_z$), but, it conserves $L_z - S_z$, where
$L_z$ is the $z$-component of the total orbital angular momentum.
The conservation of $L_z - S_z$ ensures that $V$ represents proper interactions
for the low-lying states of the system, because as found
in Ref.\onlinecite{xie}, 
there are almost degenerate lowest-energy states with the same value
of $L_z - S_z$, which are well separated from higher energy states.
Similarly, using the particle-hole symmetry, our theory can also be extended
to the case of $\nu = 1 + \epsilon$, where the wave function has $L_z + S_z$
symmetry and $Q$ changes from 1 to 0 at the transition. \cite{ahn}

To calculate the physical quantities, we approximate all the interactions
by their thermal averages except for $V$ in the Hamiltonian. 
The problem is then reduced to finding the properties of the reduced
Hamiltonian $H_{red}=\sum_{m,\sigma}\epsilon_{m,\sigma}c_{m,\sigma}^{\dagger}
c_{m,\sigma} + V$, where $\epsilon_{m,\sigma}$ is the self-energy for the 
electron in the state $| m,\sigma \rangle $.
In a mean-field concept, we ignore the fluctuations about the 
expectation values of $c_{m+1 \uparrow}^{\dagger}c_{m\downarrow}$.
This suggests that it will be useful to express such a product 
of operators formally as 
$(c_{m+1,\uparrow}^{\dagger}c_{m,\downarrow}-b_{m}) + b_{m}$, and
subsequently neglect quantities which are bilinear in the presumably small
fluctuation term in parentheses.
If we follow this procedure with our reduced Hamiltonian, we obtain
a model Hamiltonian
\begin{eqnarray}
H_{M}=\sum_{m,\sigma}\epsilon_{m,\sigma}
c_{m,\sigma}^{\dagger}c_{m,\sigma} 
\nonumber \\ 
 -\sum_{m}(\Delta_{m}^{*}
c_{m+1,\uparrow}^{\dagger}c_{m,\downarrow}+\Delta_{m}
c_{m,\downarrow}^{\dagger}c_{n+1,\uparrow}+\Delta_{m}^{*}b_{m}),
\label{hm}
\end{eqnarray}
where $b_m$'s are to be determined self-consistently from the relation
$b_{m}=\langle c_{m+1,\uparrow}^{\dagger}c_{m,\downarrow}\rangle$
and $\Delta_{m}$'s are the order parameters defined as
\begin{eqnarray}
\Delta_m = \sum_{n}V_{m,n+1,n,m+1}\langle 
c_{n+1,\uparrow}^{\dagger} c_{n,\downarrow} \rangle.
\label{order}
\end{eqnarray}
As was done by the Bogoliubov transformation in the BCS theory,
the Hamiltonian in Eq. (\ref{hm}) can be diagonalized by a suitable
linear transformation which is specified by
\begin{eqnarray}
c_{m,\downarrow}^{\dagger}&=&v_{m}\gamma_{m,0}^{\dagger}
+u_{m}^{*}\gamma_{m,1}^{\dagger}
\nonumber \\
c_{m+1,\uparrow}&=&-u_{m}^{*}\gamma_{m,0} + v_{m}\gamma_{m,1},
\label{trsf}
\end{eqnarray}
where the coefficients $u_m$ and $v_m$ satisfy $|u_m|^2+|v_m|^2=1$
and $\gamma_{m,0}$ and $\gamma_{m,1}$ are the new Fermi operators.
Using these new operators, the model Hamiltonian in Eq. (\ref{hm})
is expressed as
\begin{eqnarray}
H_{M}=-\sum_{m}\Delta_{m}b_{m}^{*}
+\sum_{m}E_{m}( \gamma_{m,0}^{\dagger} \gamma_{m,0}-
\gamma_{m,1}^{\dagger} \gamma_{m,1})
\nonumber \\ +\sum_{m} \eta_{m}
( \gamma_{m,0}^{\dagger} \gamma_{m,0}+\gamma_{m,1}^{\dagger} \gamma_{m,1}),
\label{hm2}
\end{eqnarray}
where $\eta_{m}=(\epsilon_{m+1,\uparrow}+\epsilon_{m,\downarrow})/2$,
$E_{m}=\sqrt{\xi_{m}^{2}+\Delta_{m}^{2}}$, and
$\xi_{m}= ( \epsilon_{m,\downarrow} - \epsilon_{m+1,\uparrow})/2$.
Here $E_{m}$ and $\xi_{m}$ satisfy the following relations;
\begin{eqnarray}
|v_{m}|^2=1-|u_{m}|^2=\frac{1}{2}(1-\frac{\xi_{m}}{E_{m}}).
\label{umvm}
\end{eqnarray}
If we measure the energies with respect to the chemical potential
lying between the up- and down-spin levels, we can ignore the last term
in Eq. (\ref{hm2}) because $\eta_{m}$ is negligibly small, as compared to
$E_{m}$.

The ground state $|\Psi_{G} \rangle$ of $H_{M}$ is then written as
\begin{eqnarray}
|\Psi_{G} \rangle 
= \prod_{m} \gamma_{m,1}^{\dagger}|0\rangle
= \prod_{m} (u_{m}c_{m,\downarrow}^{\dagger}+v_{m}c_{m+1,\uparrow}^{\dagger})
|0\rangle,
\label{ground}
\end{eqnarray}
where $|0 \rangle $ is the vacuum state, and the ground state energy which
is evaluated analytically \cite{fertig,ahn} is found to be lower than
those for the states with $\Delta_{m} = 0$. 
Note that the ground state is just equivalent to the skyrmion wavefunction
introduced in Ref.\onlinecite{fertig}.
This result suggests that {\it the quantum mechanical origin of the
skyrmion in the quantum Hall system is electronic pairing.}

Since the Hamiltonian $H_{M}$ in Eq. (\ref{hm2}) describes independent
excitations and the $\gamma_{m}$ operators obey Fermi statistics,
the probability of the excitation in thermal equilibrium is given by
$\langle \gamma_{m,0}^{\dagger} \gamma_{m,0} \rangle
= 1- \langle \gamma_{m,1}^{\dagger} \gamma_{m,1} \rangle
=f(E_{m})$, where $f(E_{m})$ is the usual Fermi function.
Using this relation and the linear transformation in Eq. (\ref{trsf}),
Eq. (\ref{order}) can be rewritten as 
\begin{eqnarray}
\Delta_{m}&=&\sum_{n}^{~~~\prime}V_{m,n+1,n,m+1} u_{n}^{*} v_{n}
\langle \gamma_{m,1}^{\dagger} \gamma_{m,1}
- \gamma_{m,0}^{\dagger} \gamma_{m,0} \rangle
\nonumber \\
&=&\frac{1}{2}\sum_{n}^{~~~\prime}V_{m,n+1,n,m+1}
\frac{\Delta_{n}}{E_{n}}\tanh \frac{1}{2} \beta E_{n},
\label{gapeq}
\end{eqnarray}
where $\beta=1/k_{B}T$ and the prime denotes $m \neq n$.
In the first equality in Eq. (\ref{gapeq}), the expectation values
of the off-diagonal terms such as $\gamma_{m,0}^{\dagger} \gamma_{m,1}$
are neglected because their contributions are not important in
thermal equilibrium.
By solving self-consistently the order parameter equation, 
we can obtain 
$\Delta_{m}$ as a function of $T$ for each magnetic field and find that
the order parameters have non-zero values only when $B < B_{c}$ and
$T < T_{c}$, where $B_{c}$ and $T_{c}$ are the critical magnetic field
and critical temperature, respectively.
Since there is an effective positive charge in the vicinity of the
origin, $\xi_{m}$ has in fact the dependence of $m$, where $m$'s are
small integer numbers.
For computational convenience, assuming $\xi_{m} = \xi$
for all values of $m$, 
we can determine $\xi$ in unit of the typical Coulomb energy $e^{2}/\epsilon l$
from the self-consistent equation $\xi = \tilde{g} + E_{ex}
\tanh(\beta \xi /2)$, where $\tilde{g} = \frac{1}{2} g^{*} \mu_{B}
B/(e^{2}/\epsilon l)$ is the Zeeman coupling parameter, $l=\sqrt{\hbar
c /e B_{\perp}}$ is the magnetic length, and $B$ and $B_{\perp}$ are the
total applied mgnetic field and its normal componenet to the 2D plane,
respectively.
Here $g^{*}$ and $\epsilon$ are the band effective $g$-factor and
the dielectric constant of the host material in the quantum Hall sample.
To determine the unknown parameter $E_{ex}$ that comes from the
exchange interaction, we first calculate the critical value of $\xi$
from Eq. (8) at T = 0, which gives zero values for the order parameters
$\Delta_{m}$.
Using the known critical value ($\tilde{g}_c$ = 0.0265) for the Zeeman coupling
parameter, \cite{ahn} at which the skyrmion state disappears, the value
of $E_{ex}$ is determined so that the critical value $\xi_{c}$ is equivalent
to $\tilde{g}_c + E_{ex}$ at T = 0.
Then, this value of $E_{ex}$ is used for the self-consistent solutions
of Eq. (8) at finite temperatures.
In fact, our approach of using $E_{ex}$ as a parameter and a constant value
for $\xi_{m}$ change slightly the magnitudes of $\Delta_{m}$,
however, it does not affect the main result of our calculations.

Once $\Delta_{m}(T)$ is calculated, the temperature-dependent excitation
energy $E_{m}$ and the occupation number $f_{m} = f(E_{m})$
can be determined.
Then, the specific heat is calculated by $C = T dS/dT$, where
$S$ is the electronic entropy of a fermion gas, defined as
$S = -2k_{B} \sum_{m}[(1-f_{m}) \ln (1-f_{m}) +f_{m} \ln f_{m}]$.
In the 2D electron system with the magnetic field perpendicular to the plane,
since there is no Meissner effect, the system has no latent heat and
the entropy is continuous at $T_{c}$.
Fig. \ref{fig1} shows the discontinuity of the specific heat at $T_{c}$,
and the size of the discontinuity is readily evaluated as follows;
\begin{eqnarray}
\Delta C= (C_{s}-C_{n}) \left|_{T_{c}}
= -\frac{ f(\xi_{c}) \left[ 1-f(\xi_{c}) \right]}{k_{B}T_{c}}
(\frac{-d\Delta^{2}}{dT}) \right|_{T_{c}},
\end{eqnarray}
where $\xi_{c}=\xi(T_{c})$, $\Delta^{2}= \sum_{m} |\Delta_{m}|^{2}$,
and the subscripts $s$ and $n$ denote the $skyrmion$ and $normal$ phases,
respectively.
Here we point out that for a given nonzero $\tilde{g}$, the size of
the discontinuity in the specific heat $\Delta C$ is finite for a single
skyrmion state, however, $\Delta C /C_{n}$ is negligible in the thermodynamic
limit $N \rightarrow \infty$.
Thus, practically, the observation of the phase transition will be only
possible when the Zeeman coupling is very weak so that the skyrmion size
becomes macroscopic or many skyrmions contribute to the discontinuity of
the specific heat.
In the multi-skyrmion case, when we do not take into account inter-skyrmion
interactions, the discontinuity of the specific heat at the filling factor
$\nu$ may be approximated by $\Delta C (\nu) \approx \Delta C N_{\phi}
|\nu -1 |$.  
It should be addressed that our skyrmion phase transition will not be
the second-order phase transition, because no long-range order will exist
for a finite-size skyrmion.
The skyrmion phase transition involves the localized spins of electrons
and is not related to the long-range order, similarly to the Kondo problem,
where there appears the quasi-bound state between the conduction-electron spin
and the localized impurity spin below $T_{c}$. \cite{nagaoka}

As already known for zero-temperature systems, there is a critical
magnetic field for the {\it skyrmion-normal} state transition.\cite{ahn}
Based on our pairing theory in the quantum Hall system, we can determine 
the temperature-dependent critical magnetic field $B_{c}(T)$, similarly
to the the Gorter-Casimir phenomenological theory for superconductors, which
gives the relation $B_{c}(T)/B_{c}(0) \approx 1-(T/T_{c})^{2}$.
With $B_{\perp}$ fixed, $B_{c}(T)$ in our system can be expressed as
\begin{eqnarray}
 \frac{\tilde{g_{c}}(T)}{\tilde{g_{c}}(0)}  =
 \frac{B_{c}(T)}{B_{c}(0)} 
\approx \left[ 1-\left(\frac{T}{T_{c}}\right)^{3} \right]^{1/2},
\label{bct}
\end{eqnarray}
where $k_{B} T_{c}$ is calculated to be about $0.24 e^{2}/\epsilon l$
for $\tilde{g} = 0$.
This formula for $B_{c}(T)$ is found to be in good agreement
with the results obtained directly from the order parameter
equation in Eq. (\ref{gapeq}) (see Fig. \ref{fig2}).
For GaAs samples, $T_{c}$ for $\tilde{g} \approx 0$ is approximately
$12.0 \sqrt{B_{\perp} [{\rm Tesla}]}$ K.\cite{til}
If fluctuations or impurity effects are considered,
the critical temperature is expected to be much lowered,
however, this is beyond the scope of our work.
Since fluctuations above the mean field is important in two 
dimensional systems, we expect that the sharp specific heat jump
as shown in Fig. \ref{fig1} will not take place in real systems.
The estimation of to what extent $B_{c}(T)$ and $\Delta C(T)$ are smeared
out by the fluctuations, thus, will make our work more explicit as the Ginzburg
criterion does in the usual Landau theory.
Unfortunately, we are unable to find any reliable criterion for the validity of 
our mean field theory at this stage, which will be our future work.

In summary, we have presented an analogy between the skyrmion state
and the superconducting state.
The counterpart of the Cooper pair in superconductivity is found to
be the pair which consists of a minority-spin electron and a majority-spin hole.
We find the temperature-dependent critical magnetic field $B_{c} (T)$, above
which the skyrmion is unstable at temperature $T$.
Since our approach is based on the mean field approximation and the 
fluctuations above the mean field are usually important in two dimensional
systems, our calculated $B_{c}(T)$ is considered to be an upper bound for
the true temperature-dependent critical magnetic field.

We would like to thank Dr. Y.-J. Kim for useful discussions.
This work is supported by the MOST, the CTPC and CMS at Korea Advanced
Institute of Science and Technology.

\begin{figure}
\caption{ The relative specific heat $C_{s}-C_{n}$ is drawn as a function
of T for the Zeeman coupling parameter $\tilde{g}=0.01$. }
\label{fig1}
\end{figure}

\begin{figure}
\caption{ The temperature-dependent critical magnetic field scaled
by $B_{c}(0)$ is shown as a function of $T/T_{c}$, with $B_{\perp}$ fixed.
The dotted and solid lines represent the results from the order parameter
equation and the fitting formula 1-$(T/T_{c})^3$, respectively. }
\label{fig2}
\end{figure}


\begin{references}
\bibitem{prange} {\it The Quantum Hall Effect}, edited by
R. E. Prange and S. M. Girvin (Springer-Verlag, New-York, 1990).

\bibitem{barret} S. E. Barret, G. Dabbagh, L. N. Pfeiffer, K. W. West,
and R. Tycko, Phys. Rev. Lett. {\bf 74}, 5112 (1995);
R. Tycko, S. E. Barret, G. Dabbagh, L. N. Pfeiffer, and K. W. West,
Science {\bf 268}, 1460 (1995). 

\bibitem{lee} D.-H. Lee and C. L. Kane, Phys. Rev. Lett. {\bf 64},
1313 (1990); S. L. Sondhi, A. Karlhede, S. A. Kivelson, and
E. H. Rezayi, Phys. Rev. B {\bf 47}, 16419 (1993); K. Moon, H. Mori,
K. Yang, S. M. Girvin, A. H. MacDonald, L. Zheng, D. Yoshioka, and
S.-C. Zhang, Phys. Rev. B {\bf 51}, 5138 (1995).

\bibitem{fertig} H. A. Fertig, L. Brey, R. Cote, and A. H. MacDonald,
Phys. Rev. B {\bf 50}, 11018 (1994).

\bibitem{macdonald} A. H. MacDonald, H. A. Fertig, and L. Brey,
Phys. Rev. Lett. {\bf 76}, 2153 (1996).

\bibitem{xie} X. C. Xie and S. He, Phys. Rev. B {\bf 53}, 1046 (1996).

\bibitem{ahn} K.-H. Ahn and K. J. Chang, Phys. Rev. B {\bf 55}, 6735 (1997);
Superlatt. Microstruct. (in press).

\bibitem{bcs} J. Bardeen, L. N. Cooper, and J. R. Schrieffer,
Phys. Rev. {\bf 106}, 162 (1957); {\bf 108}, 1175 (1957).

\bibitem{interaction} The single-particle eigenstates are obtained
in the symmetric guage.
$V_{m, n+1, n, m+1}$ can be computed exactly
from its analytical form. However, because of the computational difficulty
arising from numerical errors, we model $V_{m, n+1, n, m+1}$ for
$m+n >$ 50 by $a(m+n)b(|m-n|)e^{2}/\epsilon l$, where
$a(x)=(2.82580+1.03174x)^{-0.39458}$ and
$b(x)=\exp(0.05013-0.00866x^{2}-0.62085\sqrt{x})$, while
for $m+n \leq 50 $ we use the analytical formulas.

\bibitem{nagaoka} Y. Nagaoka, Phys. Rev. {\bf 138A}, 1112 (1965).

\bibitem{til}
In this case, $B_{\perp}$ is fixed with varying the total magnetic field,
which can be realized in tilted-field experiments.

\end{references}
\end{document}